\documentclass[aps,showpacs]{revtex4-2}
\usepackage{graphicx,color}
\def\bc{\begin{center}}
\def\ec{\end{center}}

\def\beq{\begin{equation}}
\def\eeq{\end{equation}}

\def\br{{\bf r}}

\def\bR{{\bf R}}
\def\bq{{\bf q}}
\def\bQ{{\bf Q}}
\def\bk{{\bf k}}

\begin{document}

\title{Quantum evolution with random phase scattering}

\author{K. Ziegler}
\affiliation{Institut f\"ur Physik, Universit\"at Augsburg\\
D-86135 Augsburg, Germany}
\date{\today}

\begin{abstract}
We consider the quantum evolution of a fermion-hole pair in a d-dimensional gas of non-interacting fermions in the presence of
random phase scattering. This system is mapped onto an effective Ising model, which enables us to show rigorously that the
probability of recombining the fermion and the hole decays exponentially with the distance of their initial spatial separation. 
In the absence of random phase scattering the recombination probability decays like a power law, which is reflected by an infinite
mean square displacement. The effective Ising model is studied within a saddle point approximation and yields a finite
mean square displacement that depends on the evolution time and on the spectral properties of the deterministic part of
the evolution operator.  
\end{abstract}

\maketitle

\tableofcontents

\newpage

\section{Introduction}

\begin{figure}[t]
\includegraphics[width=8cm,height=1.5cm]{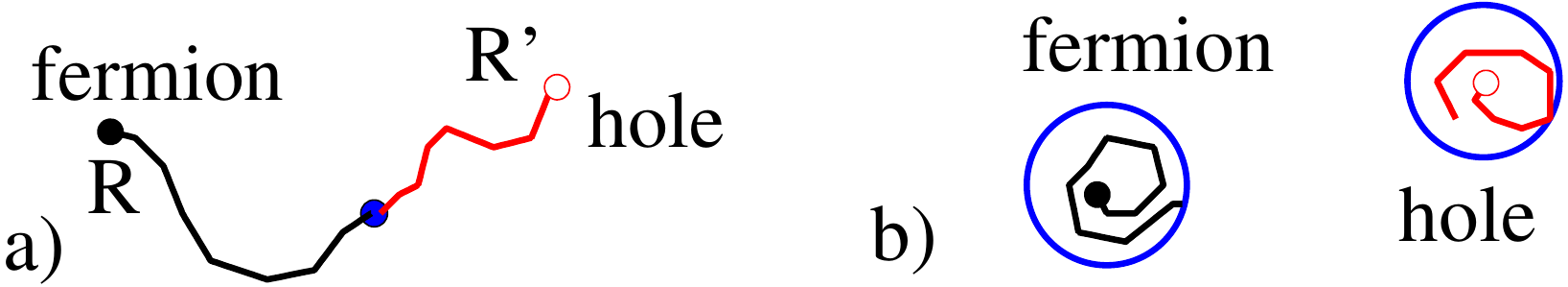}
    \caption{a) There is a recombination of a fermion created at $\bR$ and a hole created at $\bR'$ at the blue dot. 
     b) The localization of a fermion and a hole near their points of creation implies an exponentially small probability for
     recombination. The localization radius (or decay length) is indicated by the blue circles. In this sketch we consider 
     $|\bR-\bR'|$ much larger than the localization radius.
    }    
\label{fig:recomb}
\end{figure}

The creation of an electron-hole pair and its subsequent recombination is a fundamental process in quantum physics with
many applications in different fields. Although there exist phenomenological descriptions of this process by classical decay
models~\cite{PhysRevLett.10.162,Oneil1990DynamicsOE}, for a deeper understanding a quantum approach is required.
We will focus here on a fermion-hole pair in a $d$--dimensional system of non-interacting fermions. The pair can be created either by
photons and phonons in a real material or by injection into the system. Then the question is, whether this pair recombines after some 
evolution by emitting a photon/phonon or the fermion and the hole remain localized near the place where they were created 
initially (cf. Fig. \ref{fig:recomb}). Both possibilities can be studied by measuring the return probability to the initial 
quantum state. This probability depends on the spatial separation of the
fermion and the hole. Assuming that the hole is created at the site $\bR$ and the fermion at the site $\bR'$, we can define the
probability $P_{\bR\bR'}$ that the system returns to the initial state over the finite time interval $\tau$. Although it
is plausible that this probability decreases with increasing distance $|\bR-\bR'|$, the law of change with the distance
depends on the interaction with the environment. For instance, on a periodic lattice this probability has a long range
behavior, which depends on the dimensionality of the underlying space. In the following we will focus on the effect of
random phase scattering on the spatial decay of this probability. In other words, is it possible to control the spatial fermion-hole 
separation to avoid their recombination?

To analyze the evolution and calculate physical quantities, the standard procedure would be to diagonalize the Hamiltonian
$H$ of the evolution operator $e^{-iH\tau/\hbar}$. For a translational invariant system this can be achieved through a
Fourier transformation. However, in a realistic system the Hamiltonian $H$ is not translational invariant but subject to some 
disorder. In this case the corresponding random Hamiltonian cannot be diagonalized by a Fourier transformation.
To mimic the effect of disorder in the evolution of a quantum state we ``scramble'' a translational invariant 
$e^{-iH\tau/\hbar}$ 
with a random phase factor $e^{i\alpha}$ by using the evolution operator $U=e^{i\alpha}e^{-iH\tau/\hbar}$.
This choice was inspired by the random unitary gate models that have been discussed in the context of quantum 
circuits~\cite{PhysRevA.97.023604,PhysRevX.9.031009,conmatphys-031720-030658}.
The following analysis is also inspired by previous studies of the invariant measure of transport in systems
with random chiral Hamiltonians~\cite{Ziegler_2015}. Although this seems to be an entirely different problem, 
there are some striking similarities that are reflected by their graphical representations.

\section{Summary of the main results}

The central result of this work is that the random phase scattering of non-interacting fermions is equivalent to scattering on
discrete Ising spins or on a continuous (real) Ising field. This is a consequence of a geometric restriction in the graphical
representation due to the Fermi statistics. It enables us to deform the Ising field integration such that the poles of
the integrand of the return probability $P_{\bR'\bR}$ are avoided. This implies an exponential decay with respect
to $|\bR-\bR'|$. For an explicit evaluation of the decay we employ a saddle point approximation of the Ising field. 
This provides for $h=e^{-iH\tau}$
\[
P_{\bR'\bR}=\langle|(\phi+h)^{-1}_{\bR\bR'}|^2\rangle_\phi\approx |(\phi_0+h)^{-1}_{\bR-\bR'}|^2
,
\]
where $\phi_0$ is determined by a saddle point equation. The corresponding mean square displacement reads
\[
[R_\nu^2]=\frac{\tau^2}{2}\int_\bk \frac{(\partial_{k_\nu}\epsilon_\bk)^2}
{[1+\phi_0^2+2\phi_0\cos (E_0+\epsilon_\bk\tau)]^2}
\Big/\int_\bk \frac{1}{1+\phi_0^2+2\phi_0\cos (E_0+\epsilon_\bk\tau)}
,
\]
where $\epsilon_\bk$ is the dispersion of the Hamiltonian $H$ and $E_0$ is related to the Fermi energy.
In the absence of random phase scattering we have $\phi_0=1$, which implies for $E_0+\epsilon_\bk\tau\le {\bar E}$
\[
[R_\nu^2]\sim \frac{2\pi\tau}{3d\mu}(\pi-{\bar E})^{-2}
\]
when ${\bar E}<\pi$, and $[R_\nu^2]$ is infinite for ${\bar E}\ge\pi$.

\section{Quantum evolution with random phase scattering}
\label{sect:rphase2}

A system of non-interacting fermions with the Hamiltonian $H$ evolves during a fixed time step $\tau$ with the unitary 
evolution operator $U_\tau=e^{i\alpha}e^{-iH\tau}$. Here and subsequently we have chosen the scale of physical 
quantities such that $\hbar=1$.
The phases $\{\alpha_\br\}$ are randomly distributed on $[-\pi,\pi)$, independently on different lattice sites $\br$. 
$U_\tau$ acts on the $2^{|\Lambda|}$ dimensional Hilbert space, spanned by the fermionic states $|\{n_\br\}\rangle$
with occupation numbers $n_\br=0$ or $n_\br=1$ on a lattice site $\br\in\Lambda$. $|\Lambda|$ is the number of lattice sites
and $|\{n(0)\}\rangle\equiv|\{n_\br(0)\}\rangle$ is the initial state, in which the fermionic system is prepared at the beginning. 
Then we consider the situation in which a fermion-hole pair is created by the operator $c^\dagger_{\bR'}c^{}_\bR$
at time $t=0$ at different sites $\bR$, $\bR'$ . To determine the 
spatial correlation of the fermion-hole pair after the time $\tau$, where the quantum state is the initial state again, we write
\[
\frac{\langle|\langle\{n(0)|U_\tau c^\dagger_{\bR'}c^{}_\bR|\{n(0)\}\rangle|^2\rangle_\alpha}
{|\langle\{n(0)|U_\tau|\{n(0)\}\rangle|^2\rangle_\alpha}
,
\]
which is the return probability for the initial state $|\{n(0)\}\rangle$.
To avoid the specific definition of the initial state, we sum over the return probabilites of all basis states to obtain 
\beq
\label{rec_prob00a}
P_{\bR\bR'}:=\frac{\langle|Tr(U_\tau c^\dagger_{\bR'}c^{}_\bR)|^2\rangle_\alpha}{\langle|Tr U_\tau|^2\rangle_\alpha}
,
\eeq
where $Tr$ is the trace of $2^{|\Lambda|}\times 2^{|\Lambda|}$ matrices. For $\tau=0$ we have
\[
\langle\{n\}| c^\dagger_{\bR'}c^{}_\bR|\{n\}\rangle=\delta_{\bR\bR'}\delta_{n_\bR,1}
,
\]
such that
only the particle number operator $c^\dagger_\bR c^{}_\bR$ contributes to the trace, while the spatially separated 
fermion-hole pair does not. For $\tau>0$, though, the evolution $U_\tau c^\dagger_{\bR'}c^{}_\bR|\{n\}\rangle$
can create some overlap with $|\{n\}\rangle$, which contributes to the trace. Thus, the return probability
$P_{\bR\bR'}$ ($\bR'\ne\bR$) is a measure for how effective the evolution with $U_\tau$ can move the fermion-hole pair
to the same site. It is plausible that this is less likely the larger the distance $|\bR-\bR'|$ is and that it increases with
increasing time $\tau$. Therefore, $P_{\bR\bR'}$ decays with this distance and may increase with $\tau$. 

Besides creating a fermion and a hole simultaneously, we can also create a hole at site $\bR$ and time $t=0$, let this hole
evolve for the time $\tau$ and then annihilate it. The probability for this annihilation process reads
\beq
\label{rec_prob00b}
P'_{\bR\bR'}:=\frac{\langle|Tr(c^\dagger_{\bR'}U_\tau c^{}_\bR)|^2\rangle_\alpha}{\langle|Tr U_\tau|^2\rangle_\alpha}
.
\eeq
In the following we drop the index $\tau$ for simplicity and use $U_\tau\equiv U$.
Assuming that $|\{{\tilde n}_\bq\}\rangle\equiv|\{{\tilde n}\}\rangle$ are eigenstates of $U$ for a special realization of the phases
$\{\alpha_\br\}$, we obtain
\beq
\label{corr02a}
{\tilde C}_{\bQ,t}:=
Tr (U^{(1-t)}c^\dagger_\bQ U^tc^{}_{\bQ})
=\sum_{\{{\tilde n}_\bq\},\{{\tilde n}'_\bq\}}\langle\{{\tilde n}\}| U^{(1-t)}\{{\tilde n}\}\rangle\langle\{{\tilde n}\}|c^\dagger_\bQ|
\{{\tilde n}'\}\rangle  \langle\{{\tilde n}'\}|
U^t|\{{\tilde n}'\}\rangle\langle\{{\tilde n}'\}|c^{}_{\bQ}|\{{\tilde n}\}\rangle
\eeq
for $t=0,1$. With 
$\langle\{{\tilde n}\}|c^\dagger_\bQ|\{{\tilde n}'\}\rangle\langle\{{\tilde n}'\}|c^{}_{\bQ}|\{{\tilde n}\}\rangle
=\delta_{{\tilde n}'_{\bQ},0}\delta_{{\tilde n}_{\bQ},1}\prod_{\bq\ne\bQ}\delta_{{\tilde n}'_\bq,{\tilde n}_\bq}$ 
we get
\[
{\tilde C}_{\bQ,t}=\sum_{\{{\tilde n}_\bq\},\{{\tilde n}'_\bq\}}\langle\{{\tilde n}\}| U^{(1-t)}|\{{\tilde n}\}\rangle
\langle\{{\tilde n}'\}|U^t|\{{\tilde n}'\}\rangle
\delta_{{\tilde n}'_{\bQ},0}\delta_{{\tilde n}_{\bQ},1}\prod_{\bq\ne\bQ}\delta_{{\tilde n}'_\bq,{\tilde n}_\bq}
.
\]
Since $U$ is diagonal in this basis with $\langle\{{\tilde n}\}| U|\{{\tilde n}\}\rangle
=\prod_\bq\langle {\tilde n}_\bq|e^{-iE_\bq {\tilde n}_\bq}|{\tilde n}_\bq\rangle
=\prod_\bq e^{-iE_\bq {\tilde n}_\bq}$, we obtain a product of diagonal matrix elements
\[
\langle\{{\tilde n}\}| U^{(1-t)}|\{{\tilde n}\}\rangle\langle\{{\tilde n}'\}|U^t|\{{\tilde n}'\}\rangle
=\prod_{\bq}e^{-iE_\bq {\tilde n}_\bq (1-t)}e^{-iE_\bq {\tilde n}'_\bq t}
.
\]
Inserting this into Eq. (\ref{corr02a}), we get for the sum due to the Kronecker deltas
\beq
\label{corr03a}
{\tilde C}_{\bQ,t}
=Tr (U^{(1-t)}c^\dagger_\bQ U^tc^{}_\bQ)
=e^{-iE_\bQ(1-t)}\prod_{\bq\ne\bQ}(1+e^{-iE_\bq})
=\frac{e^{iE_\bQ t}}{1+e^{iE_\bQ}}\prod_{\bq}(1+e^{-iE_\bq})
.
\eeq
Finally,  we return to the real-space representation to obtain $e^{-iE_\bq}\to \hat{U}_{\br\br'}$, where $\hat{U}$ is a 
$|\Lambda|\times|\Lambda|$ matrix on the lattice, and
\beq
\label{correlator02}
Tr (Uc^\dagger_{\bR'}c^{}_\bR)=({\bf 1}+\hat{U}^\dagger)^{-1}_{\bR\bR'}\det({\bf 1}+\hat{U})
\ ,\ \ 
Tr (c^\dagger_{\bR'}Uc^{}_\bR)=({\bf 1}+\hat{U})^{-1}_{\bR\bR'}\det({\bf 1}+\hat{U})
,
\eeq
where $det$ is the corresponding determinant. Hence the return probabilities become
\beq
\label{rec_prob_c}
P_{\bR\bR'}
=\frac{\langle|({\bf 1}+\hat{U}^\dagger)^{-1}_{\bR\bR'}\det({\bf 1}+\hat{U})|^2\rangle_\alpha}
{\langle|\det({\bf 1}+\hat{U})|^2\rangle_\alpha}
\ ,\ \ 
P'_{\bR\bR'}
=\frac{\langle|({\bf 1}+\hat{U})^{-1}_{\bR\bR'}\det({\bf 1}+\hat{U})|^2\rangle_\alpha}
{\langle|\det({\bf 1}+\hat{U})|^2\rangle_\alpha}
\eeq
due to Eqs. (\ref{rec_prob00a}), (\ref{rec_prob00b}). The identity $|({\bf 1}+\hat{U}^\dagger)^{-1}_{\bR\bR'}|^2
=|({\bf 1}+\hat{U})^{-1}_{\bR'\bR}|^2$ implies that $P'_{\bR\bR'}=P_{\bR'\bR}$.

\section{Functional integral representation}
\label{sect:random_phase_1}

For the further treatment of the return probability $P'_{\bR\bR'}=P_{\bR'\bR}$  in Eq. (\ref{rec_prob_c}) it is convenient
to separate the random phase factor and the deterministic evolution of $\hat{U}$ as $\hat{U}_{\br\br'}=e^{i\alpha_\br}h_{\br\br'}$. Then we employ a Grassmann functional integral to write
\beq
\label{fi001}
P_{\bR'\bR}=\frac{1}{\cal N}\langle\int_\varphi\exp\left[
\pmatrix{
\varphi_1\cr
\varphi_2\cr
}\cdot\pmatrix{
{\bf 1}+e^{i\alpha}h & 0\cr
0 & {\bf 1}+h^\dagger e^{-i\alpha}\cr
}\pmatrix{
\varphi'_1\cr
\varphi'_2\cr
}\right]\varphi_{1\bR}\varphi'_{1\bR'}\varphi_{2\bR'}\varphi'_{2\bR}\rangle_\alpha
\]
\[
=\frac{1}{\cal N}\langle adj_{\bR\bR'}({\bf 1}+e^{i\alpha}h)adj_{\bR'\bR}({\bf 1}+h^\dagger e^{-i\alpha})\rangle_\alpha
=\frac{1}{\cal N}\langle|\det({\bf 1}+e^{i\alpha}h)|^2|({\bf 1}+e^{i\alpha} h)^{-1}_{\bR\bR'}|^2\rangle_\alpha
\eeq
with the normalization
\[
{\cal N}=\langle\int_\varphi\exp\left[
\pmatrix{
\varphi_1\cr
\varphi_2\cr
}\cdot\pmatrix{
{\bf 1}+e^{i\alpha}h & 0\cr
0 & {\bf 1}+h^\dagger e^{-i\alpha}\cr
}\pmatrix{
\varphi'_1\cr
\varphi'_2\cr
}\right]\rangle_\alpha
=\langle|\det({\bf 1}+e^{i\alpha}h)|^2\rangle_\alpha
.
\]
We note that the kernel of the quadratic form has zero modes (i.e., eigenmodes of ${\bf 1}+e^{\alpha}h$ with vanishing eigenvalue)  because the eigenvalues of the random unitary matrices $e^{i\alpha}h$  and
$h^\dagger e^{-i\alpha}$ are randomly distributed on the unit circle in the complex plane. These zero modes depend on the
realization of the random phase.

In the integral (\ref{fi001}) we pull out the phase factors by rescaling the Grassmann fields to obtain
\beq
\label{fi002}
P_{\bR'\bR}=\frac{1}{\cal N}
\langle\int_\varphi\exp\left[
\pmatrix{
\varphi_1\cr
\varphi_2\cr
}\cdot\pmatrix{
e^{-i\alpha}+h & 0\cr
0 & e^{i\alpha}+h^\dagger\cr
}\pmatrix{
\varphi'_1\cr
\varphi'_2\cr
}\right]\varphi_{1\bR}\varphi'_{1\bR'}\varphi_{2\bR'}\varphi'_{2\bR}\rangle_\alpha
\]
\[
=\frac{1}{\cal N}\int_\varphi
\prod_\br\langle(1+e^{-i\alpha_\br}\varphi_{1\br}\varphi'_{1\br})(1+e^{i\alpha_\br}\varphi_{2\br}\varphi'_{2\br})\rangle_\alpha
\exp\left[
\pmatrix{
\varphi_1\cr
\varphi_2\cr
}\cdot\pmatrix{
h & 0\cr
0 & h^\dagger\cr
}\pmatrix{
\varphi'_1\cr
\varphi'_2\cr
}\right]\varphi_{1\bR}\varphi'_{1\bR'}\varphi_{2\bR'}\varphi'_{2\bR}\rangle_\alpha
,
\]
which gives after phase averaging
\[
=\frac{1}{\cal N}\int_\varphi
\prod_\br(1+\varphi_{1\br}\varphi'_{1\br}\varphi_{2\br}\varphi'_{2\br})
\exp\left[
\pmatrix{
\varphi_1\cr
\varphi_2\cr
}\cdot\pmatrix{
h & 0\cr
0 & h^\dagger\cr
}\pmatrix{
\varphi'_1\cr
\varphi'_2\cr
}\right]\varphi_{1\bR}\varphi'_{1\bR'}\varphi_{2\bR'}\varphi'_{2\bR}
.
\eeq 
We get the same result when we replace the phase factor by an Ising spin $\{S_\br=\pm1\}$ or by a real Gaussian field $\phi_\br$
which will be called Ising field in the following.
For the latter we write
\beq
\label{fi005}
P_{\bR'\bR}=
\frac{1}{{\cal N}_\phi}\int e^{-\frac{1}{2}\sum_\br\phi_\br^2}\int_\varphi
\exp\left[
\pmatrix{
\varphi_1\cr
\varphi_2\cr
}\cdot\pmatrix{
\phi+h & 0\cr
0 & \phi+h^\dagger\cr
}\pmatrix{
\varphi'_1\cr
\varphi'_2\cr
}\right]\varphi_{1\bR}\varphi'_{1\bR'}\varphi_{2\bR'}\varphi'_{2\bR}\prod_\br d\phi_\br
\]
\[
=\frac{1}{{\cal N}_\phi}\int e^{-\frac{1}{2}\sum_\br\phi_\br^2}adj_{\bR\bR'}(\phi+h)adj_{\bR'\bR}(S+h^\dagger)\prod_\br d\phi_\br
=\frac{1}{{\cal N}_\phi}\int e^{-\frac{1}{2}\sum_\br\phi_\br^2}|adj_{\bR\bR'}(\phi+h)|^2\prod_\br d\phi_\br
\]
\[
=\langle|(\phi+h)^{-1}_{\bR\bR'}|^2\rangle_\phi
\ \ {\rm with}\ \ \ \langle \ldots\rangle_\phi
=\frac{1}{{\cal N}_\phi}\int e^{-\frac{1}{2}\sum_\br\phi_\br^2}|\det(\phi+h)|^2\ldots\prod_\br d\phi_\br
.
\eeq
This result is reminiscent of the average two-particle Green's function with respect to a Gaussian distribution of $\phi_\br$, 
multiplied by the determinant term $|\det(\phi+h)|^2$. There are two important differences in comparison to the average
two-particle Green's function $\langle|(V+H_0)^{-1}_{\bR\bR'}|^2$ of Anderson localization, though. The first is that the determinant
can be written as a product of the eigenvalues of $\phi+h$. This cancels poles of $|(\phi+h)^{-1}_{\bR\bR'}|^2$, implying
that the poles of the Green's functions are not relevant for the $\phi_\br$ integration. In other words, the adjugate matrix
$adj_{\bR\bR'}(\phi+h)=\det(\phi+h)(\phi+h)^{-1}_{\bR\bR'}$ does not have any pole for $|\phi_\br|<\infty$, and the 
integration with respect to $\phi_\br$ can be deformed in any finite area of the complex plane. This reflects an exponential
decay with a finite decay length of the return probability $P_{\bR'\bR}$. The second difference is that $h$ is unitary and
its eigenvalues are located 
on the unit circle of the complex plane, while the Hamiltonian $H_0$ in the Anderson localization problem is a Hermitian matrix 
with eigenvalues on the real axis.

The deformation of the $\phi_\br$ integration provides a rigorous but only qualitative result regarding the decay of $P_{\bR'\bR}$.
For a quantitative result of the decay we must perform the integration explicitly. We will do that approximately within
a saddle-point integration in Sect. \ref{sect:spa} and App. \ref{app:spi}.

\section{Discussion} 
\label{sect:Discussion}

\begin{figure}[t]
\includegraphics[width=7.5cm,height=3cm]{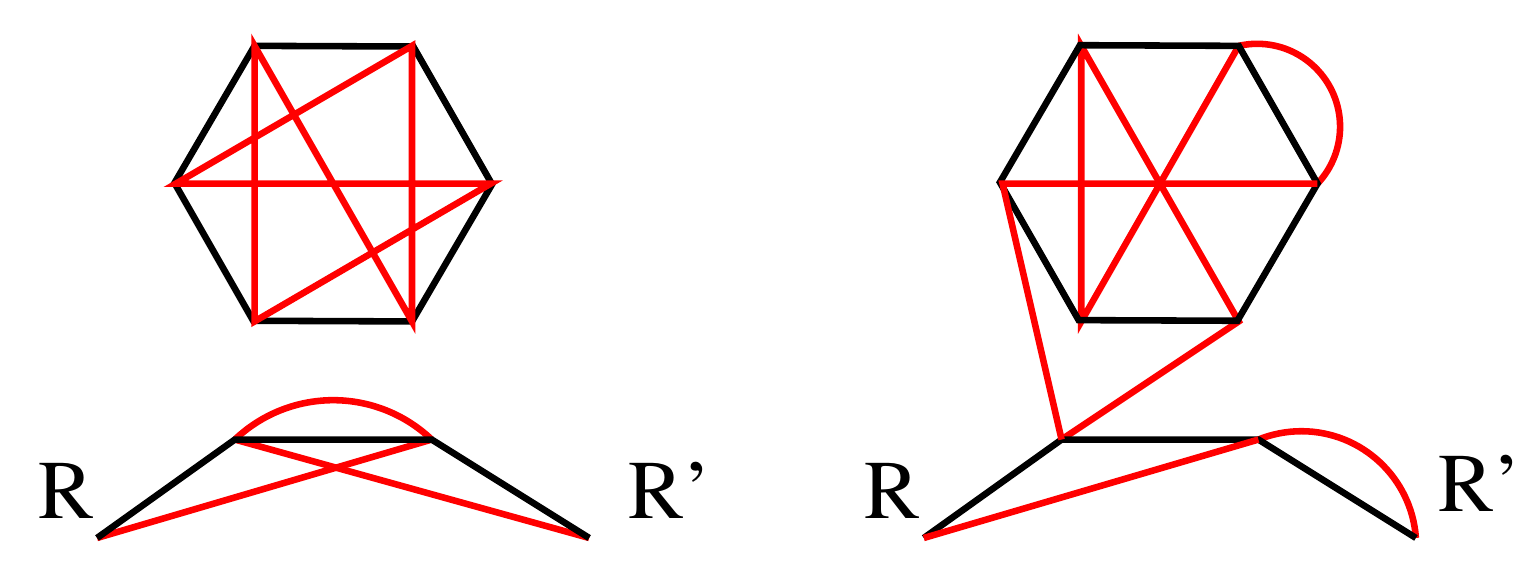}
    \caption{Two typical graphs representing contributions to the functional integral of the return probability $P_{\bR'\bR}$ in
     Eq. (\ref{fi001}).  Black (red) edges represent $h$ ($h^\dagger$). Both edges form (i) a loop and (ii) a string connecting
    the sites $\bR$ and $\bR'$ of the fermion and the hole. The strings are contributions to the inverse matrix elements
    $({\bf 1}+e^{i\alpha}h)^{-1}_{\bR\bR'}$ and $({\bf 1}+h^\dagger e^{-i\alpha})^{-1}_{\bR\bR'}$, respectively,
    while loops are contributions to the determinants. There are only 4-vertices, except for the endpoints
    $\bR$ and $\bR'$, which are connecting to two black and two red edges. Other edge crossings are not connected by vertices.
    }    
\label{fig:graphs}
\end{figure}

First, we note that the expansion of the integrand of Eq. (\ref{fi001}) and a subsequent Grassmann and phase integration
yields graphs with 4-vertices, where two edges $h_{\br\br'}$ 
from $\varphi_1$ and two Hermitean conjugate edges $h^\dagger_{\br\br'}$ from $\varphi_2$ are connected.
This condition is enforced by the Grassmann property, which requires a product of 
$\varphi_{1\br}\varphi'_{1\br}\varphi_{2\br}\varphi'_{2\br}$ at each site $\br$. Moreover, the random phase factors
glue $h$ and $h^\dagger$ at these products to form a 4-vertex and to prevent a 2-vertex. The geometric property of 
the 4-vertex enables us either to form loops of edges or to connect the sites $\bR$ and $\bR'$ by a string of both types of edges.
Both, the $h$ edges as well as the $h^\dagger$ edges form loops and an $\bR$-$\bR'$ string separately. This is a consequence
of the diagonal kernel of the quadratic form in Eq.  (\ref{fi001}). Moreover, each loop carries a factor $-1$ from the
Grassmann field.
Two typical examples are depicted in Fig. \ref{fig:graphs} with the same formation of the nine black edges but with different 
formations of the nine red edges.
In the left example a loop and a double string are separated by a special choice of red edges, while in the right example there is 
only one connected graph.


This type of graphs is known from 
the invariant measure of chiral random Hamiltonians~\cite{Ziegler_2015}. There is a crucial difference though that
is related to the zero mode: In contrast to the random phase scattering $e^{i\alpha_\br}h_{\br\br'}$, the scattering 
of the chiral model is $e^{i\alpha_\br}h_{\br\br'}\sum_{\br''}h^\dagger_{\br'\br''}e^{-i\alpha_{\br''}}$. For the latter
we have a uniform zero mode
\beq
\sum_{\br'} [\delta_{\br\br'}-e^{i\alpha_\br}h_{\br\br'}\sum_{\br''}h^\dagger_{\br'\br''}e^{-i\alpha_{\br''}}]=0
\eeq
for any realization of the random phase.

\subsection{Hopping expansion}
\label{sect:hopping_expansion}

In order to get a better understanding of the behavior of the return probability $P_{\bR'\bR}$ we return to the expression of 
Eq. (\ref{rec_prob_c}) with random phases and simplify it by neglecting the determinants. This leads to the product of
the conjugate one-particle Green's functions of only two individual particles:
\[
\langle({\bf 1}+he^{i\alpha})^{-1}_{\bR\bR'}({\bf 1}+e^{-i\alpha}h^\dagger)^{-1}_{\bR'\bR}\rangle_\alpha
=\langle(e^{-i\alpha}+h)^{-1}_{\bR\bR'}(e^{i\alpha}+h^\dagger)^{-1}_{\bR'\bR}\rangle_\alpha
.
\]
A hopping expansion of the inverse matrices in powers of the evolution operator $e^{i\alpha}h$ and its Hermitian conjugate
can be written as a truncated geometric series
\[
({\bf 1}+e^{i\alpha}h)^{-1}_{\bR\bR'}({\bf 1}+h^\dagger e^{-i\alpha})^{-1}_{\bR'\bR}
=\sum_{l,m=0}^{N-1}(e^{i\alpha}h)^l_{\bR\bR'}(h^\dagger e^{-i\alpha})^m_{\bR'\bR}
,
\]
where the truncation with $N<\infty$ is necessary because it is not clear whether the series converges. 
Since after phase averaging only $l=m$ survives, we can ignore terms with $l\ne m$ here. This gives
\beq
\label{series01a}
\sum_{l=0}^{N-1}(he^{i\alpha})^l_{\bR\bR'}(e^{-i\alpha}h^\dagger)^l_{\bR'\bR}
=\delta_{\bR\bR'}+h_{\bR\bR'}h^\dagger_{\bR'\bR}+\sum_{\br_1,\br_1'}h_{\bR\br_1}h_{\br_1\bR'}
h^\dagger_{\bR'\br_1'}h^\dagger_{\br_1'\bR}e^{i\alpha_{\br_1}-i\alpha_{\br_1'}}
\]
\[
+\ldots +\sum_{\br_1,\br_1',\br_2,\br_2',\ldots,\br_{N-1},\br_{N-1}'}
h_{\bR\br_1}h_{\br_1\br_2}\cdots h_{\br_{N-1}\bR'}
h^\dagger_{\bR';\br_{N-1}'}h^\dagger_{\br_{N-1}'\br_{N-2}'}\cdots h^\dagger_{\br_1'\bR}\prod_{j=1}^{N-1}
e^{i\alpha_{\br_j}-i\alpha_{\br_j'}}
.
\eeq
Now we can average over the random phases to obtain
\beq
\label{av_series01a}
\langle({\bf 1}+he^{i\alpha})^{-1}_{\bR\bR'}({\bf 1}+e^{-i\alpha}h^\dagger)^{-1}_{\bR'\bR}\rangle_\alpha
=\delta_{\bR\bR'}+h_{\bR\bR'}h^\dagger_{\bR'\bR}+\sum_{\br_1}h_{\bR\br_1}h_{\br_1\bR'}
h^\dagger_{\bR'\br_1}h^\dagger_{\br_1\bR}
\]
\[
+\ldots +\sum_{\br_1,\br_2,\ldots,\br_{N-1}}\sum_{\pi_{N-1}}
h_{\bR\br_1}h_{\br_1\br_2}\cdots h_{\br_{N-1}\bR'}
h^\dagger_{\bR';\pi(\br_{N-1})}h^\dagger_{\pi(\br_{N-1})\pi(\br_{N-2})}\cdots h^\dagger_{\pi(\br_1)\bR}
,
\eeq
where we sum with respect to all permutations $\pi_{N-1}$ of all non-degenerate sites of $\{\br_1,\br_2,\ldots,\br_{N-1}\}$. 
Although this is a compact
expression, it is difficult to perform the sum over the permutations and to calculate the corresponding values. 
Nevertheless, as an important special case the identity $\pi_{N-1}=id$ can be calculated. It is a
contribution of an unrestricted random walk on the lattice. This represents a long range correlation in the form of diffusion. 
However, it will be destroyed by the determinant factor in Eq. (\ref{fi001}), as mentioned in the previous section, where
the Ising field representation leads to an exponential decay. In other words, the coupling of many fermions to the random
phase scattering supports localization by avoiding singularities that appear in the case of two particles. For a quantitative
result of the exponential decay we study the mean square displacement within a saddle point approximation in the next section.

\subsection{Saddle point approximation}
\label{sect:spa}

The return probability in Eq. (\ref{fi005}) is treated within the saddle point integration of the Ising field $\phi$ 
(cf. App. \ref{app:spi}). This yields
\beq
P_{\bR'\bR}=\langle|(\phi+h)^{-1}_{\bR\bR'}|^2\rangle_\phi\approx |(\phi_0+h)^{-1}_{\bR-\bR'}|^2
,
\eeq
where we have neglected the flucuations $\delta\phi_\br$ around the saddle point $\phi_0$. This approximation enables us to
factorize the return probability as
\[
P_{\bR'\bR}\approx |C_{\bR-\bR'}|^2
\ ,\ \ 
C_{\bR-\bR'}=(\phi_0+h)^{-1}_{\bR-\bR'}
,
\]
where $C_{\bR-\bR'}$ can be represented by its Fourier transform 
\beq
\label{inverse_m1}
{\tilde C}_\bk=\frac{1}{\phi_0+e^{-iE_\bk}}
\eeq
with the eigenvalue $E_\bk$ of the translational invariant matrix $H\tau$. Thus, the effect of the random phase scattering
is associated only with the value of $\phi_0$, where the latter is determined by $E_\bk$ via the saddle point equation
(\ref{spa_condition}). Moreover, $\phi_0=1$ represents the absence of random phase scattering. 

The results for the Ising field $\phi_0$ of App. \ref{app:spi} can be interpreted in terms of the magnetic properties
of the classical Ising model~\cite{itzykson1989statistical}. The asymmetric shift $\cos E_0$ plays the role of an 
external magnetic field and $\phi_0$ corresponds to the magnetization~\cite{McCoyWu+1973}.
Thus, the effective Ising model has a unique Ising field $\phi_0>0$ or $\phi_0<0$ when $\cos E_0\ne 0$,
while for $\cos E_0=0$ there are either two degenerate solutions with opposite signs of $\phi_0$ (ferromagnetic phase)
or a single solution with $\phi_0=0$ (paramagnetic phase). In contrast to the classical Ising model with a continuous
transition though, Fig. \ref{fig:symmetric_energy} indicates a jump of $\phi_0$ for our effective Ising model.

\subsection{Mean square displacement}

\begin{figure}
a)
\includegraphics[width=7cm,height=7cm]{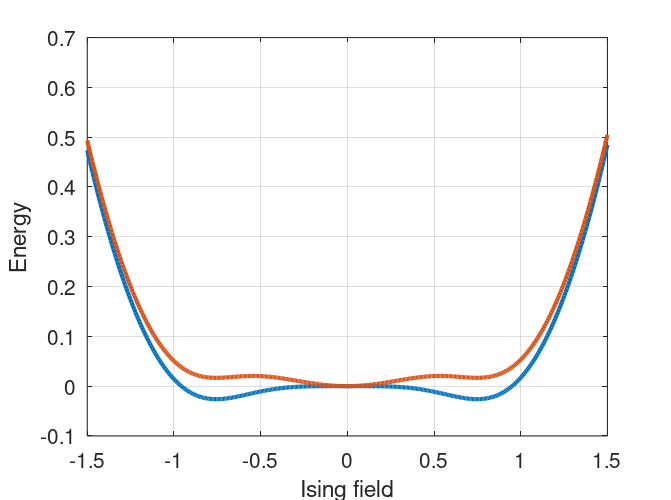}
b)
\includegraphics[width=7cm,height=7cm]{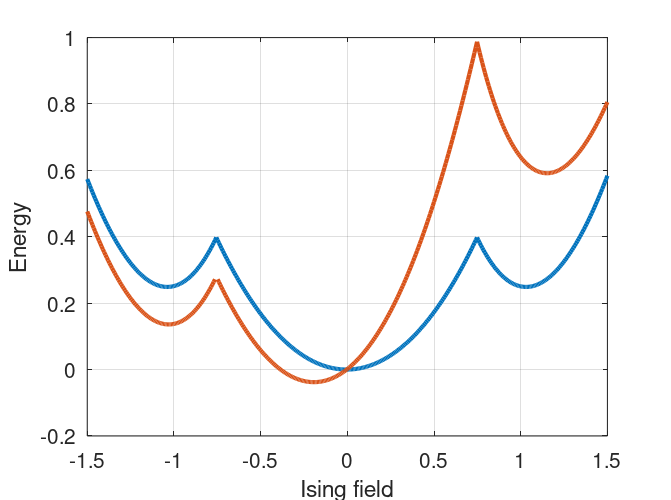}
    \caption{
    The Ising energy ${\cal E}(\phi)$ as defined in Eq. (\ref{Ising_energy1}) for a constant density of states.
    a) ${\cal E}(\phi)$ is plotted at the symmetry point $E_0=\pi/2$ for the band width $a=1.9$ (blue curve) and $a=2.1$ (red curve).
    It indicates a jump of the Ising field from two degenerate nonzero values to $\phi_{0;min}=0$. 
    b) ${\cal E}(\phi)$ is plotted with band width     $a=7.5$ for a symmetric band $E_0=0$ (red curve) and a band that
    is shifted by $E_0=\pi/2$ (blue curve). 
    }    
\label{fig:symmetric_energy}
\end{figure}

The mean square displacement provides a measure for the localization length. It is defined as
\beq
\label{msd}
[R_\nu^2]:=\frac{\sum_{\bR'}(\bR_\nu-\bR'_\nu)^2P_{\bR'\bR}}{\sum_{\bR}P_{\bR'\bR}}
=\frac{-\partial_{q_\nu}^2{\tilde P}_\bq\Big|_{\bq=0}}{{\tilde P}_0}
,
\eeq
where ${\tilde P}_\bq$ is the Fourier transform of the translational-invariant $P_{\bR'\bR}\equiv P_{\bR'-\bR}$.
Now we study $P_{\bR\bR'}=|C_{\bR\bR'}|^2$ with the help of the saddle point integration. In this case the
mean square displacement reads 
\beq
\label{msd1}
[R_\nu^2]=\frac{\sum_\bR R_\nu^2|C_\bR|^2}{\sum_\bR|C_\bR|^2}
,
\eeq
where the numerator is
\[
\sum_\bR R_\nu^2e^{i\bq\cdot\bR}|C_\bR|^2\Big|_{\bq=0}
=-\partial^2_{q_\nu}\sum_\bR e^{i\bq\cdot\bR}|C_\bR|^2\Big|_{\bq=0}
=-\partial^2_{q_\nu}\sum_\bR\int_\bk\int_{\bk'} e^{i(\bq-\bk-\bk')\cdot\bR}{\tilde C}_\bk{\tilde C}^*_{\bk'}\Big|_{\bq=0}
\]
with the Fourier transform ${\tilde C}_\bk$ of Eq. (\ref{inverse_m1}). Then the $\bR$ summation can be performed and leads to 
a Kronecker delta, which gives for  Eq. (\ref{msd1})
\[
[R_\nu^2]=-\partial^2_{q_\nu}\int_\bk {\tilde C}_\bk{\tilde C}^*_{\bq-\bk}\Big|_{\bq=0}
\Big/\int_\bk{\tilde C}_\bk{\tilde C}^*_{-\bk}
.
\]
Now we assume that $C_{-\bk}=C_{\bk}$ to obtain eventually
\beq
[R_\nu^2]=\int_\bk |\partial_{k_\nu}{\tilde C}_\bk|^2\Big/\int_\bk|{\tilde C}_\bk|^2
,
\eeq
which becomes with Eq. (\ref{inverse_m1})
\beq
\label{msd_spa}
[R_\nu^2]=\int_\bk \frac{(\partial_{k_\nu}E_\bk)^2}{(1+\phi_0^2+2\phi_0\cos E_\bk)^2}
\Big/\int_\bk \frac{1}{1+\phi_0^2+2\phi_0\cos E_\bk}
.
\eeq
A special case is the one in which the energy in Eq. (\ref{Ising_energy0}) is symmetric with respect to $\phi_0\to-\phi_0$.
Then there exists a critical value $\tau_c$ of the evolution time $\tau$:
If $\tau$ exceeds $\tau_c$ the saddle point is always $\phi_{0;min}=0$, as indicated in Fig. \ref{fig:symmetric_energy}a).
This implies for Eq. (\ref{inverse_m1}) that ${\tilde C}_\bk= e^{iE_\bk}$, which yields for the corresponding mean square
displacement
\beq
[R_\nu^2]=\int_\bk(\partial_{k_\nu}E_\bk)^2=\tau^2\int_\bk(\partial_{k_\nu}\epsilon_\bk)^2
.
\eeq
$\epsilon_\bk$ is the dispersion of the Hamiltonian $H$. Thus, the mean square displacement increases with the
squared evolution time $\tau$.
For $\tau<\tau_c$, on the other hand, or when the energy is asymmetric with respect to $\phi_0\to-\phi_0$, we have $\phi_0\ne0$.

In the absence of random phase scattering we have $\phi_0=1$ directly from Eq. (\ref{fi001}).
Then for the special case $E_\bk=k^2\tau/2\mu$ ($0\le k\le \lambda$) the mean square displacement on a $d$-dimensionsional lattice
reads
\beq
[R_\nu^2]
=\frac{\tau}{d\mu}\int_0^{{\bar E}}\frac{E^{d/2}}{(1+\cos E)^2}dE\Big/\int_0^{{\bar E}}\frac{E^{d/2-1}}{1+\cos E}dE
\eeq
with the integration cut-off ${\bar E}=\lambda^2\tau/2\mu$. This is a finite expression for ${\bar E}<\pi$, which 
diverges with a power law as
\beq
[R_\nu^2]\sim \frac{2\pi\tau}{3d\mu}(\pi-{\bar E})^{-2}
\eeq
when we approach ${\bar E}=\pi$ from below. For ${\bar E}\ge\pi$ the mean square displacement is always infinite without
random phase scattering. This result has the form of a diffusion relation with time $\tau$ and a divergent diffusion coefficient for
${\bar E}\to \pi$ when we ignore the fact that ${\bar E}$ also depends on $\tau$. A possible interpretation is that the fermion-hole
pair is subject to diffusion due to its interaction with the other fermions of the system. The divergence, on the other hand,
reflects a long range correlation of the fermion and the hole that reflects the pole of ${\tilde C}_\bk=1/(1+e^{-iE_\bk})$. 

A more detailed analysis, especially for the evaluation of $\tau_c$, requires specific expressions of the dispersion 
$\epsilon_\bk$. This would exceed the goal of this work to present a generic approach for the effect of disorder on the
recombination of fermion-hole pairs. 

\section{Conclusions and outlook}

The probability $P_{\bR'\bR}$, which describes the probability to return to the initial quantum state after the creation of
a fermion at site $\bR$ and a hole at site $\bR'$ and their evolution, decays always exponentially with the distance 
$|\bR-\bR'|$ in the presence of random phase scattering. To obtain this rigorous result a mapping of the random phase model onto 
an Ising-like model was essential.
This was supplemented by an approximative calculation of this decay, based on a saddle point integration of the
effective Ising model, to get some quantitative insight into the decay. The latter calculation is instructive, since it demonstrates 
how the solution of the saddle point equation avoids 
the singularities of the underlying fermion model. In the absence of random phase scattering, one of these singularities 
leads to a non-exponential decay for a sufficiently long evolution of the state with the fermion-hole pair. This is reflected by an
infinite mean square displacement of the fermion and the hole.
 
In our approximation we have not included the Gaussian fluctuations around the saddle point solution.
It would be interesting to include them and to determine their effect on the decay of the return probability.
In this context it would also be useful to understand the effect of these fluctuations on the transition from $\phi_0\ne 0$
to $\phi_0=0$ at the symmetry point under an increasing evolution time. Another extension of our approach is the
application to the return probability of a system under periodically repeated projective
measurements~\cite{Gruenbaum2013,PhysRevE.95.032141} 
or under randomly repeated projective measurements~\cite{Ziegler_2021}. 
Then the effect of random phase scattering on the resulting monitored evolution could also be described by the effective Ising
field model. Even more interesting but also more challenging would be the extension of the approach to the 
transition probability for the monitored evolution under randomly repeated projective 
measurements~\cite{PhysRevA.103.022222}.

\appendix

\section{Saddle point integration}
\label{app:spi}

We approximate the integral
\[
\langle \ldots\rangle_\phi
=\frac{1}{{\cal N}_\phi}\int e^{-\frac{1}{2}\sum_\br\phi_\br^2}|\det(\phi+h)|^2\ldots\prod_\br d\phi_\br
\]
by using a saddle-point integration. Then we determine the maximal contribution to the integrand by assuming a uniform
$\phi$ and write $\phi_\br=\phi+\delta\phi_\br$. This enables us to approximate the integral in terms of
Gaussian fluctuation around the uniform $\phi$ with respect to $\delta\phi_\br$, where $\phi$ must be fixed as $\phi_0$
at the minimum of the Ising energy
\beq
\label{Ising_energy0}
{\cal E}(\phi)=\frac{1}{2}\phi^2-\int_{-\infty}^\infty\log(1+\phi^2+2\phi\cos E)\rho(E)dE
\eeq
with the density of states $\rho(E)$.  The integrand is singular for $\phi=1$, $E=\pi$ and for $\phi=-1$, $E=0$. 
These singularities yield  large values for the energy. Therefore, they do not represent lowest energy contributions of the
saddle point. This is also reflected in the curves of Fig. \ref{fig:symmetric_energy}b).
Moreover, the saddle-point solution $\phi_0$ must satisfy
\beq
\label{spa_condition}
{\cal E}'(\phi)=\phi-2\int_{-\infty}^\infty\frac{\phi+\cos E}{1+\phi^2+2\phi\cos E}\rho(E)dE=0
.
\eeq
For a constant density of states $\rho(E)$ on the interval $[-a/2+E_0,a/2+E_0]$ we get
\beq
\label{Ising_energy1}
{\cal E}(\phi)=\frac{1}{2}\phi^2-\frac{1}{a}\int_{-a/2+E_0}^{a/2+E_0}\log(1+\phi^2+2\phi\cos E)dE
.
\eeq
A special case is $E_0=\pi/2$, where we have
\[
{\cal E}(\phi)=\frac{1}{2}\phi^2-\frac{1}{a}\int_{-a/2}^{a/2}\log(1+\phi^2-2\phi\sin E)dE
\]
with the symmetry relation ${\cal E}(\phi)={\cal E}(-\phi)$. This would also hold when the density of states is symmetric
with respect to $E=\pi/2$ in Eq. (\ref{Ising_energy0}).
The Ising energy is plotted for several values of $E_0$ and the band width $a$ in Fig. \ref{fig:symmetric_energy}.

%

\end{document}